\documentclass[11pt,eps]{article}
\usepackage{fullpage}
\usepackage{epsfig} 
\usepackage{graphicx}
\newtheorem{definition}{Definition}
\newtheorem{theorem}{Theorem}

\newtheorem{corollary}{Corollary}

\newcommand{\qed}{\hfill \mbox{\raggedright \rule{.07in}{.1in}}}

\newenvironment{proof}{\vspace{1ex}\noindent{\bf Proof}\hspace{0.5em}}
	{\hfill\qed\vspace{1ex}}

\title{An Optimal Trade-off between Content Freshness and Refresh Cost}
\author{\it Yibei Ling$^\natural$  and Jie Mi$^+$\\
$^\natural$Applied Research Laboratories, 
Telcordia Technologies,  NJ 07960\\
$^+$Department of Statistics, 
Florida International University, Miami, FL 33199
}
\date{2004}
\begin {document}
\maketitle
\begin{abstract}
Caching is an effective mechanism for 
reducing bandwidth usage and alleviating server load.  
However, 
the use of caching entails a compromise
between content freshness and refresh cost. 
An excessive refresh allows 
a high degree of content freshness at a greater cost of system resource.
Conversely, a deficient refresh inhibits
content freshness but saves the cost of resource usages. 
To address the freshness-cost problem,
we formulate the refresh scheduling problem with a generic cost model and 
use this cost model to determine an
optimal refresh frequency that gives the best
tradeoff between refresh cost and content freshness.
We prove the existence and uniqueness of an optimal refresh frequency 
under the assumptions that the arrival of content update is Poisson and
the age-related cost monotonically increases with decreasing freshness.
In addition, we provide an analytic 
comparison of system performance 
under fixed refresh scheduling and random  refresh scheduling,
showing that with the same average refresh frequency  
two refresh schedulings are mathematically equivalent in 
terms of the long-run average cost.    
\end{abstract}

\section{Introduction}
The timely information 
dissemination  
is the fundamental driving force that spurs 
ever-growing technology advancement and development.
The widespread of Web technologies makes the Internet
a de facto channel for mass distribution of 
information. 
Nowaday, popular web sites such as {\it www.cnn.com} and {\it www.msn.com}
can receive ten millions 
requests per day 
\cite{CnnCase,Wessels2001}
with normal request rate of $12,000$ per minute, and 
with peak rate of more than
$33,000$ per minute during breaking news. 
Such high demands
pose a significant overhead on both serving 
servers and networks surrounding 
the serving servers \cite{Lee2002}.
A variety of approaches and system architectures  
have been introduced to 
enable efficient content distribution while alleviating 
system load and bandwidth consumption.

\begin{figure}[htb]
\centerline{\psfig{figure=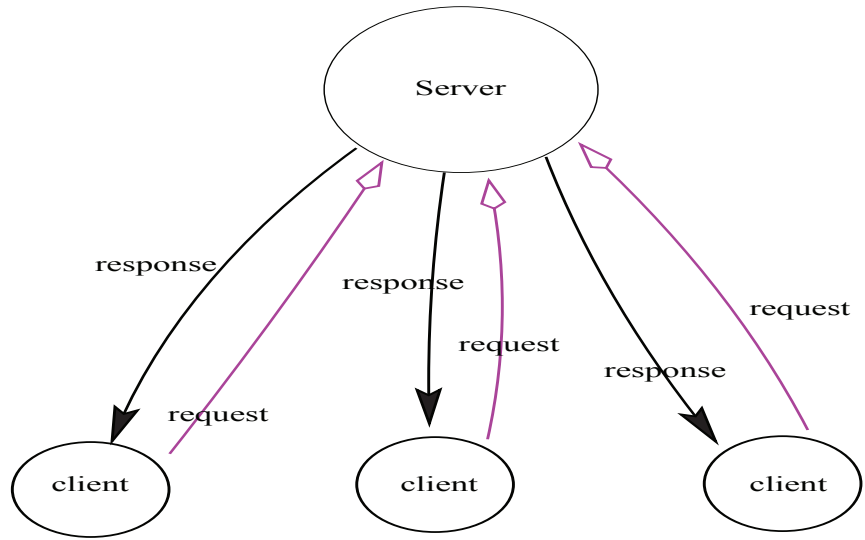,height=2.8in,width=4.5in}}
\caption{Processing flow between clients and a server}
\label{fig:figure001}
\end{figure}

\begin{figure}[htb]
\centerline{\psfig{figure=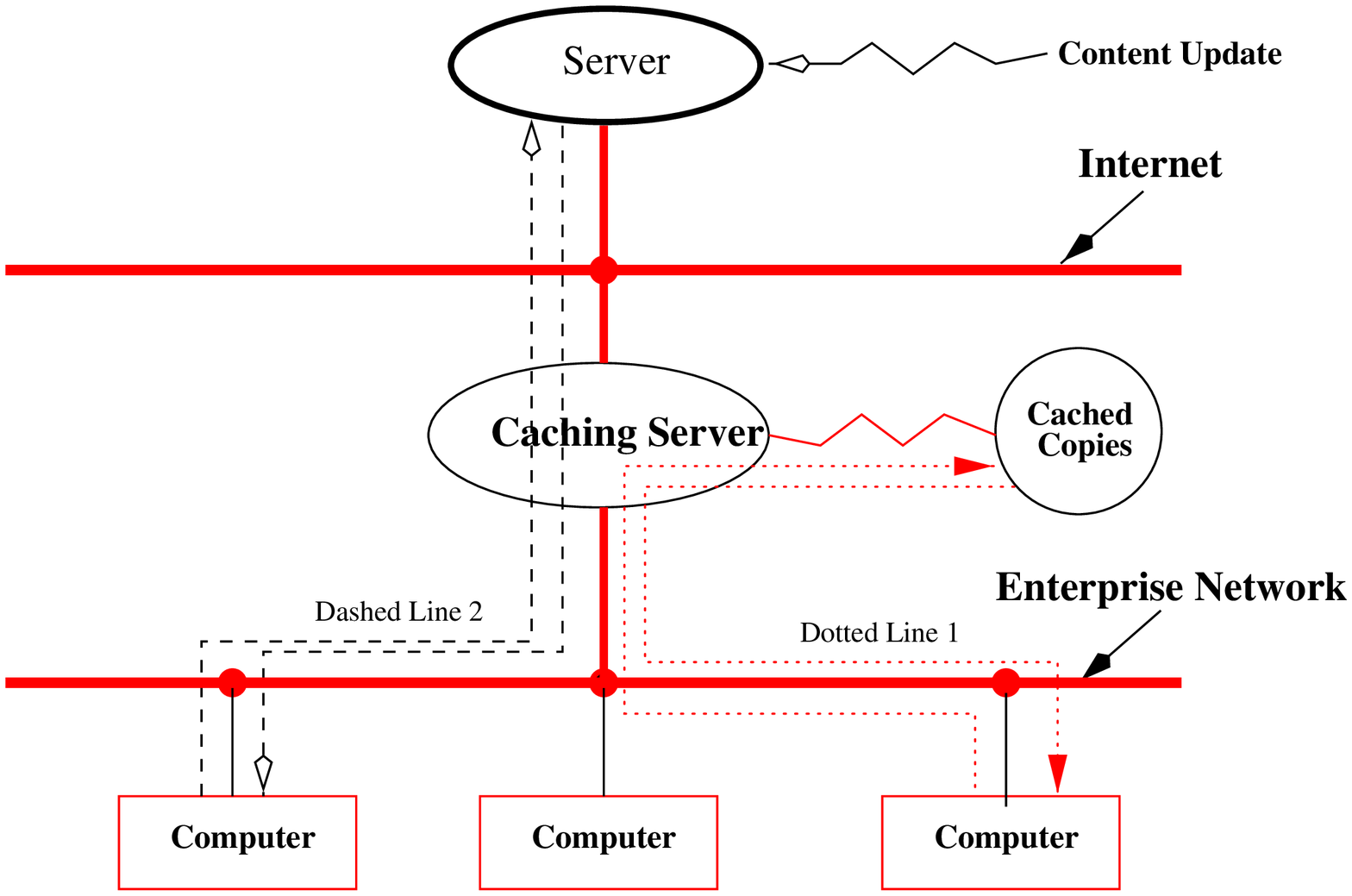,height=2.8in,width=4.5in}}
\caption{Placement of caching server at an enterprise environment}
\label{fig:figure01}
\end{figure}

Web servers and browsers (web clients) are 
the fundamental architectural building blocks 
in the World Wide Web.
A Web client is a requester of data (content) and a Web server 
is the provider of data (content).     
A web server manages and provides data source while Web
browsers send requests to a Web server for a specific source data
by means of URL (uniform resource locator).  
Upon receipt of  
a request initiated by a web client, 
the web server then processes the request and sends a response back 
to the web client. Figure~\ref{fig:figure001} illustrates 
the typical data flow between the clients and the server. 

The content resources at servers are autonomous:
they are updated independently at various rates 
without pushing updates to the clients.  
As a result, each client has to poll the remote resources 
at a web server periodically
in order to detect changes and update 
its contents. This process is referred to as 
{\it refresh synchronization}. 
The freshness comes at a cost of resource usage;   
each request initiated by a client incurs certain 
communication and computation overhead  
for processing the request.   
As a result, it could be very costly in terms of overall 
bandwidth usage and server load
when considering million of individual clients.

Caching is an effective means of reducing the system load of servers
and bandwidth usage.
The idea behind caching is to store recently 
retrieved copies of remote source 
somewhere between the clients and the remote web servers. 
As a result, the request initiated by a web client can be diverted to 
the cached copies
which are much closer to the Web client than the remote web server in terms of 
network distance.

The caching architecture in 
a representative enterprise environment is illustrated in 
Figure~\ref{fig:figure01} wherein a
caching server is placed at the enterprise's network entrance to 
external networks, acting as an intermediary between host computers (clients) inside 
the enterprise network and the internet.    
Each host machine is connecting to 
the enterprise's network backbone, 
the caching server rests between the enterprise's network
backbone and the Internet. 
Upon receiving a request originated from a host inside 
the enterprise's network,
the caching server checks to see whether the corresponding response has 
already been cached, 	
if the cached file is present, 
then the caching server returns the client with 
the cached copy, saving the client from retrieving 
the same resource (document) repeatedly from the remote server. 
The flow is represented by the dotted line 1 in Figure~\ref{fig:figure01}. 
If the cached file is absent, 
the caching server then forwards the request to a server. 
After the caching server has received 
the response from the remote server, it
returns the response to the client and locally stores  
the response for 
subsequent requests. Its 
is represented by the dashed line 2 in 
Figure~\ref{fig:figure01}. 
It is clear that the use of caching server in an enterprise 
environment enables substantial bandwidth saving and  dramatically improves 
user-perceived response time, because the cached copies are 
located within the vicinity of the clients in terms of network distance. 
The performance gain appears to be proportional to the number of users 
\cite{Cohen2001,CohenKaplan2001,Duska1997}.  

The performance gains of caching 
carry the cost of content freshness.
The cached copies immediately become obsolete as soon as the 
original copies at the remote server are updated or changed.
There is a substantial tradeoff between  content freshness and 
refresh cost:
a frequent refresh ensures the freshness of content
at high cost of
refresh. Conversely,  an 
infrequent refresh inhibits the freshness of content but saves the refresh cost.   

\begin{figure}[t]
\centerline{\psfig{file=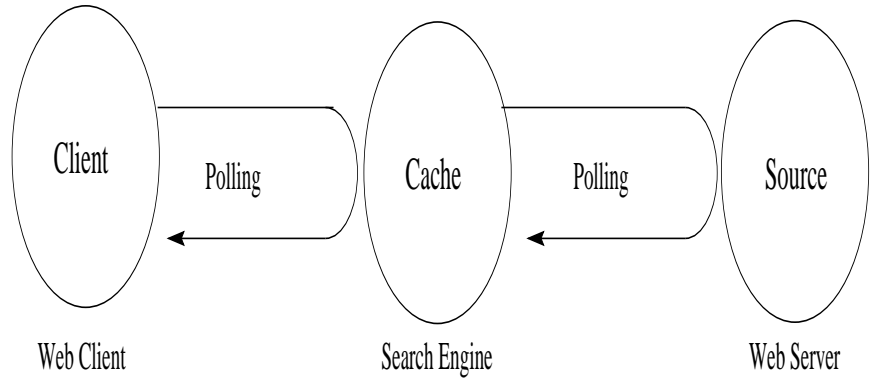,height=2in,width=4.5in}}
\caption{Conceptual Flow Diagram of Crawler-based Search Engine}
\label{fig:figure02}
\end{figure}

The freshness-cost problem also arises in crawler-based 
search engine applications. 
Crawler-based search engines such as {\it google}
provide a powerful tool for searching 
web documents, serving as a ``yellow book'' on the Web.
Figure~\ref{fig:figure02} presents a conceptual flow diagram of 
crawler-based Web search engines. Periodically, 
a Web search engine polls Web servers 
independently, 
and 
Web clients poll randomly the Web search engine searching for
the directories of Web documents. 
It is clear the refresh scheduling of the web search engine 
is independent of the
Web access by Web clients. 
In addition,
the content at Web servers may change over time and 
it is impossible to 
know a priori exactly the 
arrival time of content update.
To maintain the freshness of content, the Web search engine needs
to poll web sites and update its database directory in a frequent fashion.
Such a process is known as {\it Web crawling}. 
The freshness of content
is regarded as one of the important performance metrics for 
a crawler-based search engine \cite{Notess2001}.  
Web crawling is a prohibitively expensive computation task when considering an 
astronomical numbers of constantly changing web pages.  
It is reported that most search engines refresh their entire directory databases
once a month \cite{Notess2001}.

Cho and Garcia-Molina \cite{Cho99,Cho2000} first present a probability model 
to study the impact of various refresh (synchronization) policies on 
the content freshness with an emphasis on synchronization-order policy,
under different contexts from this paper. 
Our study in this paper differs from theirs principally
in that we consider the problem of refresh scheduling, with the objective
of optimally balancing the tradeoff between content freshness and refresh cost. 
 
The remainder of the paper is organized as follows: 
In {\it Section 2} we consider the problem of refresh scheduling
involving one cache element. 
We study and establish a generic cost 
model that accounts for both the content freshness and the refresh cost
and prove that the mathematical 
equivalence between the refresh schedulings with the 
fixed interval and the  random interval in terms of the overall refresh cost. 
{\it Section 3} 
extends the obtained results into the cases involving more than one cache elements,
with an emphasis on the {\it uniform allocation policy}.   
{\it Section 4}
concludes the paper.

\section{Mathematical Formulation and Main Results}
In this section we consider the problem of optimal refresh 
scheduling involving only one 
cache element. We begin with an introduction of relevant notions and 
definitions, followed by a cost analysis of 
the relationship between the refresh interval and the freshness of content.
Finally we identify 
the optimal refresh frequency that gives the best tradeoff between 
the content freshness and the refresh cost. 
 
We formally give the notion of 
the aggregated age function. Our definition is 
in spirit similar to the one proposed by
Cho and Garcia-Molina \cite{Cho2000,Cho99}, but differs in the sense that 
we take the ``aggregated effect" of content decay into account.
We then introduce the age-related cost function
that generalizes the notion of the aggregated age function. 

Suppose that the arrival of content update at a server follows
the Poisson process  with intensity rate 
of $\lambda$.  Let $\{X_i, i\geq 1\}$ be the interarrival times of the Poisson 
process. 
Define $S_0=0,~S_n=\sum\limits_{i=1}^n X_i$, where $S_i$ 
represents the time of the {\it i}th occurrence of 
content update at the server. Let 
\begin{equation} \label{equ:equ00}
N(t)=\sup \{n \geq 0:~S_n\leq t \}. 
\end{equation}
$N(t)$ is a random variable that represents the number of
arrivals of content updates at the server in 
the time interval $(0,t]$. 
Two closely related but different notions are given as follows.

\begin{definition} \label{def:definition1}
Under refresh frequency of $1/T$,
the age of the element $e$ 
with respect to 
the {\it i}th occurrence of content update $S_i$ at time $t \in [0,T)$ is 
$$
Age(e, S_i, t) =  (t- S_i) I_{\{t>S_i\}},
$$
\end{definition} 
where $I_{\{t>S_i\}}$ is an indicator function.  \\

$Age(e, S_i, t)$ is a function representing
a measure for the content freshness of 
the element $e$ with respect to the {\it i}th occurrence of content 
update at the server. 

\begin{definition}\label{def:definition2}
Under refresh frequency of $1/T$, the aggregated age of the 
element $e$  with respect to 
the occurrences of content update at time $t \in [0,T)$ is 
$$
A(e, t) =
\sum\limits_{i=1}^{N(t)} (t- S_i) I_{\{t>S_i\}}.$$
\end{definition} 

$A(e,t)$ is an aggregated age function that
reflects the additive effect of multiple content updates taking place 
within 
the interval $[0,t)$. 
Cho and Garcia-Molina \cite{Cho99,Cho2000} propose 
the {\it age} metric as a measure for content freshness
by only considering 
the first occurrence of content update, that is,
$Age(e,S_1,t)$. 
The major 
difference between ours and the definition by 
Cho and Garcia-Molina is that we consider the additive property of 
content freshness with respect to multiple content updates.  

\begin{figure}[ht]
\centerline{\psfig{file=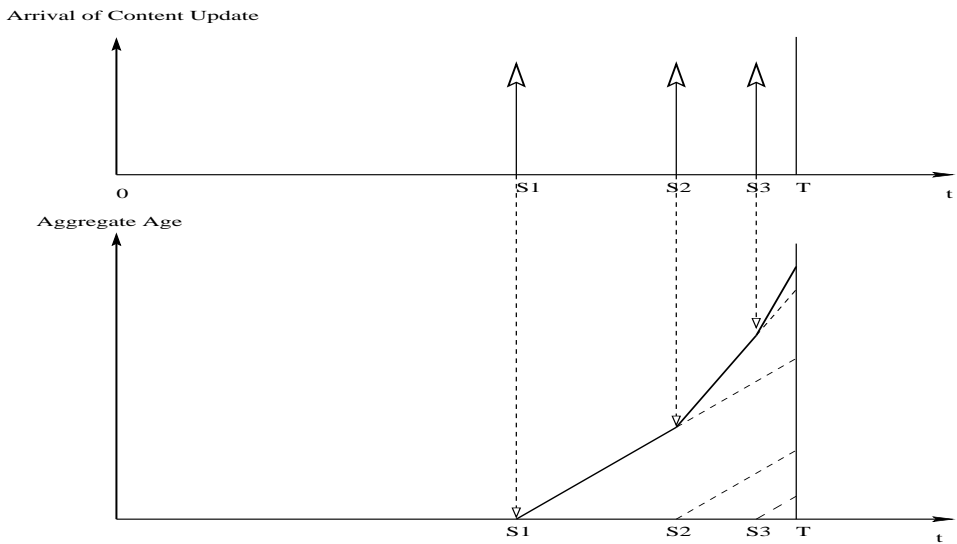,height=2.8in,width=5in}}
\caption{Evolution of Aggregated Age Function of $A(e,t)$}
\label{fig:figure002}
\end{figure}

Figure~\ref{fig:figure002} is an illustration of the evolution of the 
functions of 
the $A(e,t)$, reflecting that 
the aggregated age of the element $e$ with respect to 
the occurrences of content update over time.
 
To study the freshness-cost tradeoff, 
we introduce the age-related cost function denoted by $C_a(x)$ 
which is a nondecreasing and positive 
function of the age,  
where the variable $x$ denotes the age of the element of interest and 
the subscript $a$ the age of content (see Definition~1).
The notion of age-related cost is a generalization of the notion of 
age. As a result, the age function defined in \cite{Cho99} is 
a special case of the age-related cost function with $C_a(x)=x$.

The association of cost with freshness in the problem formulation
is partly motivated by its market relevance that
many crawler-based search 
engines such as {\it google}, {\it Inktomi} and 
{\it fast} \cite{Notess2001} introduce   
paid inclusion programs that trade the freshness of 
content and visibility for a payment.
We further assume that the cost of a refresh synchronization $C_r$, 
where the subscript $r$ indicates the cost associated with refresh.
This cost could be a measure of bandwidth usage or latency or 
a financial payment, depending on 
the choice of performance metric. 
The following theorem shows that under mild conditions 
the optimal refresh interval, $T^*$, that minimizes 
the long-run mean average cost, exists and is unique.

\begin{theorem}\label{theo:relationship}
Suppose that the arrival of update is a Poisson process with intensity 
rate $\lambda$.
Let the refresh interval be $T$, and the cost of a refresh be  $C_r$.  
Then the long-run mean average cost of caching under 
the refresh frequency of $1/T$ is given as 
\begin{equation} \label{equ:equc} 
C(T)= \frac{C_r}{T} + \frac {\lambda \int^T_0  C_a(t)dt}{T}.
\end{equation} 
If 
the age-related cost function $C_a(t)$ satisfies (1) $C_a(t)$ is differentiable and
$C^{\prime}_a(t) >0$; (2)
$C_a(\infty) = \infty$, then 
there exists a unique refresh interval, designated $T^*$, which minimizes 
the long-run mean average cost  
function $C(T)$, and $T^*$ is determined by 
the equation
$$TC_a(T)-\int\limits^T_0 C_a(t)dt=\frac{C_r}{\lambda }.$$
\end{theorem}
\begin{proof} 
For a given interval $(0,t]$, the mean average cost 
over the interval $(0,t]$ can be written as  
\begin{equation}\label{equ:cost11}
\frac{E (\hbox{the~random~cost~in~interval~}(0,t])}{t}. 
\end{equation}
Hence, the long-run mean average cost under the refresh interval $T$ is  
\begin{equation}\label{equ:cost1}
C(T)=\lim\limits_{t\to\infty} \frac{E (\hbox{the~random~cost~in~interval~}(0,t])}{t} 
\end{equation}
if this limit exists. 
Denote the random cost on the interval $((k-1)T, kT]$ by  $\xi_k(T), k \geq 1$.
Due to the properties of stationary and independent increments of 
the Poisson process \cite{Ross1996}, $\xi_k(T), k \geq 1$ are {\it iid}, 
the long-run mean average cost can be written as 
\begin{equation}\label{equ:equcost}
C(T) =\lim\limits_{t\to\infty} 
\frac{E(\lfloor \frac{t}{T} \rfloor)
\xi_1(T))}{t} 
= \frac{E(\xi_1(T))}{T},
\end{equation}
where the term $\lfloor x \rfloor$ is the floor function that gives 
the largest integer less than or equal to $x$. 

The random cost $\xi_1(T)$ on the interval $(0,T]$ 
consists of a refresh cost and the age-related cost accumulated 
over the interval $(0,T]$.
For the arrival of the {\it n}th content update at time $S_n \leq T $, 
the age-related cost, $C_a(T-S_n)$, is a function of the age $T-S_n$.
The aggregated age-related cost in the interval $(0,T]$ is thus
expressed as
$$ 
\sum\limits^{N(T)}_{n=1} \left ( C_a(T-S_n) \right )I_{\{N(T)>0\}}.
$$  
Hence, the random cost $\xi_1(T)$ on the interval $(0,T]$
is given as 
\begin{equation} \label{equ:equlimit}
\xi_1 (T) 
= C_r + \sum\limits^{N(T)}_{n=1} \left ( C_a(T-S_n) \right )I_{\{N(T)>0\}}.
\end{equation}
Note that
\begin{equation} \label{equ:equim}
\sum \limits^{N(T)}_{n=1} C_a(T-S_n)I_{\{N(T)>0\}} =  
\sum\limits^\infty _{n=1} 
C_a(T-S_n) I_{\{n \leq N(T)\}} = \sum\limits^\infty _{n=1} 
C_a(T-S_n) I_{\{S_n \leq T\}} 
\end{equation}
\noindent
and
\begin{equation} \label{equ:equi}
E\left ( C_a(T-S_n)I_{\{S_n\leq T\}}\right ) = \int\limits^T_0 C_a(T-t)f_n(t)dt, 
\end{equation}
\noindent
where $f_n(t)$ is the probability density function of $S_n$, 
and follows
the gamma distribution as below
\begin{equation}\label{equ:gamma}
f_n(t) = \frac{ \lambda^n}{(n-1)!} t^{n-1}e^{-\lambda t},~t>0.
\end{equation}
\noindent
Substituting Eq(\ref{equ:gamma}) into Eq(\ref{equ:equi}) we obtain
\begin{equation} \label{equ:equ12}
 E \left( C_a (T-S_n) I_{\{S_n\leq T\}}\right ) = \int ^T_0 C_a (T-t) 
\frac {\lambda ^n}{(n-1)!} t^{n-1} e^{-\lambda t}dt.
\end{equation}
\noindent
We then have
\begin{eqnarray} \label{equ:equ10}
E \left ( \sum\limits ^{N(T)}_{n=1} C_a(T-S_n) 
I_{\{N(T)>0\}}\right ) 
&= & \sum\limits ^\infty _{n=1} \int ^T_0 C_a(T-t) \frac {\lambda ^n}{(n-1)!}t^{n-1}e^{-\lambda t}dt \nonumber \\ 
&=& \int^T_0 C_a(T-t) \lambda e^{-\lambda t} \left (\sum \limits ^\infty_{n=1} 
\frac {(\lambda t)^{n-1}}{(n-1)!}\right ) dt \nonumber \\
&=& \lambda \int ^T_0 C_a(T-t) dt = \lambda \int ^T_0 C_a(t) dt .
\end{eqnarray} 
Therefore, by Eqs(\ref{equ:equcost}), (\ref{equ:equlimit}), 
and (\ref{equ:equ10}), the long-run mean 
average cost can be expressed as 
$$C(T)=\frac{E(\xi_1 (T))}{T}= 
\frac{C_r}{T} + \frac {\lambda \int^T_0 C_a(t)dt}{T}.$$

\noindent The derivative of   
$C(T)$ is then given as
$$C^\prime(T) = - \frac{C_r}{T^2} + \frac{\lambda C_a(T)}{T} - \frac{\lambda 
\int^T_0 C_a(t)dt}{T^2}. $$
\noindent Define a function $\varphi(T)$ as 
\begin{equation} \label{equ:varphi}
\varphi (T)\equiv T^2 C^\prime (T) = -C_r+\lambda TC_a(T)-\lambda \int^T_0 C_a(t)dt 
\end{equation}
Observe that  $C^\prime(T)$ and $\varphi(T)$ have the same sign. 
It can also be verified that the
$$\varphi ^\prime (T) = \lambda TC_a^\prime(T)>0, ~~\forall~T>0$$
\noindent 
since $C_a^\prime(t) >0, \forall t>0$.  
This shows that the function
$\varphi (T)$ strictly increases in $T>0$.  
Observe that $\varphi (0)=-C_r<0$.  
Moreover, for any fixed $\epsilon >0$, if 
$T> \epsilon$, then   
\begin{eqnarray}
\varphi(T)  & = & -C_r + \lambda \int ^T_0 \left ( C_a(T) - C_a(t)\right ) dt 
  \geq  -C_r + \lambda \int^{\epsilon}_0 (C_a(T) -C_a(t))dt \nonumber \\ 
& \geq &
 -C_r+ \lambda (C_a(T)-C_a(\epsilon))\epsilon  
\end{eqnarray}
Hence $\varphi (\infty ) \equiv \lim\limits 
_{T\to\infty} \varphi (T)=\infty$ since $C_a(\infty) = \infty$.  
Considering the facts 
that $\varphi (0) <0,~\varphi (\infty)=\infty$ and $\varphi (T)$ 
is strictly increasing in $T>0$, we conclude that there must exist 
a unique $0<T^*<\infty$ such that 
\[ C^\prime(T)= \left \{ \begin{array}{ll}
<0, & \hbox{ if } 0<T<T^* \\
 = 0, & \hbox{ if } T=T^* \\
>0, & \hbox{ if } T > T^* .
\end{array}
\right . \]
\noindent
Therefore, $T^*$ minimizes $C(T)$, or
$$T^*=\arg \left ( \min\limits _{T>0} C(T)\right ).$$
\end{proof}
  
\begin{corollary} \label{coro:coro1}
Suppose that the age-related cost function $C_a(t) = Ct$, where $C>0$ is 
the 
proportionality constant.  
Then the long-run mean average cost function $C(T)$ is given as 
\begin{equation}
C(T)=\frac{C_r}{T} + \frac {C\lambda T}{2}
\end{equation}
and is minimized at 
\begin{equation}\label{equ:cost}
T^* = \sqrt{ \frac{2C_r}{\lambda C}}.
\end{equation}
Also, $C(T^*)=\sqrt{2\lambda CC_r}$.
\noindent
\end{corollary}

\begin{proof}  Obviously $C_a(t)=Ct$ satisfies the condition 
in Theorem~\ref{theo:relationship}. 
Thus, $T^* \equiv \arg (\min\limits _{T>0} C(T))$ 
exists and is uniquely determined by the equation $C^\prime (T) =0$ 
where
$$C^\prime (T)=-\frac{C_r}{T^2} + \frac{\lambda C}{2}$$
\noindent
and the unique solution to $C^\prime (T)=0$ is given by 
$$T^*=\sqrt{ \frac{2C_r}{\lambda C}}$$. 
\end{proof} 

From Corollary~\ref{coro:coro1}, it is shown that 
the optimal refresh interval decreases with increasing rate of 
the content update when the age function is linear. 
This result, however, is also true in general as shown in the following theorem.

\begin{theorem} \label{theorem:the2} Under the same conditions of 
Theorem~\ref{theo:relationship}, 
the optimal refresh interval $T^*$ strictly decreases in $\lambda >0$. 
\end{theorem}
\begin{proof}
Clearly $T^*$ is a function of $\lambda$. For the sake of simplicity, 
we will not express this dependence explicitly.  
For the same reason we will suppress the symbol $*$ and just use 
$T$ to denote the optimal refresh interval in the following derivation. 
From Theorem~\ref{theo:relationship},  the optimal refresh 
interval satisfies $\varphi (T)=0$, or
\begin{equation} \label{equ:equ13}
-C_r + \lambda TC_a(T)-\lambda \int^T_0 C_a(t)dt=0.
\end{equation}
\noindent
Taking derivative with respect to $\lambda$ in Eq(\ref{equ:equ13}), 
we obtain
\begin{equation} \label{equ:equ14}
TC_a(T) + \lambda TC_a^{\prime}(T)\frac{dT}{d\lambda} -
\int^T_0 C_a(t)dt=0.
\end{equation}
\noindent From Eq(\ref{equ:equ14}) it follows that
\begin{eqnarray} \label{equ:equ141}
\lambda T C_a^\prime (T) \frac{dT}{d \lambda}  & = & 
\int^T_0 C_a(t)dt - TC_a(T) \nonumber \\
&=& \int^T_0 \left ( C_a(t)-C_a(T)\right ) dt <0. 
\end{eqnarray}
\noindent
Eq(\ref{equ:equ141}) immediately implies
$$\frac{dT^*(\lambda)}{d\lambda} <0$$
\noindent
since $C_a^\prime(T) >0$.  
This proves that $T^*=T^*(\lambda )$ is a 
strictly decreasing function of $\lambda$. 
\end{proof}

\begin{theorem} \label{theorem:the3}  
Under the same conditions of Theorem~\ref{theo:relationship}, 
the fixed optimal refresh policy $T^*$ strictly increases in $C_r >0$. 
\end{theorem}
The proof is simple and straightforward, therefore is omitted. \\

\noindent {\bf Remark 1}: Theorems~\ref{theorem:the2},~\ref{theorem:the3} 
demonstrate that 
the impact of the arrival rate of content update as well as the
refresh overhead $C_r$ on the optimal refresh frequency.  
A high frequency of content update and cheap 
refresh cost exact a high frequency of refresh in order to 
maintain a certain level of content freshness.
Conversely, an infrequent content update 
and expensive refresh cost require 
refresh be performed at low frequency in order to save bandwidth usage and
mitigate server load. 
This analytical results obtained above not only agree well with our intuition, 
but also provide us 
with a quantitative connection between optimal refresh interval and 
arrival rate of update.

\begin{theorem} \label{theorem:tt}
Suppose that two age-related cost functions $C_{a1}(t)$ and $C_{a2}(t)$ 
satisfy the conditions in Theorem~\ref{theo:relationship}.  
Let $T{}^*_1$ and $T{}^*_2$ be the optimal refresh intervals
associated with $C_{a1}(t)$ and $C_{a2}(t)$, respectively. 
For given $C_r$ and $\lambda$,  
if $0\leq \Delta (t) \equiv C_{a2}(t)-C_{a1}(t)$ is 
nondecreasing in $t\geq 0$, then $T{}^*_1 \geq T{}^*_2$.
\end{theorem}

\begin{proof}  
According to the proof of Theorem~\ref{theo:relationship}, 
it is shown that
\begin{equation}
\varphi (T)\equiv -C_r +\lambda TC_{a1}(T)
-\lambda \int^T_0 C_{a1}(t)dt 
\end{equation}
\noindent
strictly increases in $T$, and 
\begin{equation}
\varphi (T{}^*_1) = -C_r + \lambda T{}^*_1 C_{a1}(T{}^*_1) -
\lambda \int^{T_1^*}_0 C_{a1}(t)dt=0.
\end{equation}
\noindent
Suppose that contrary to the claimed result, i.e.,  
$T{}^*_1 < T{}^*_2$.  
From the monotonicity of 
$\varphi (T)$ in Eq(\ref{equ:varphi}) and $\varphi (T{}^*_1)=0$, 
we have $\varphi (T{}^*_2)>0$.  That is
\begin{equation}
-C_r+ \lambda T{}^*_2 C_{a1}(T{}^*_2 ) - 
\lambda \int ^{T{}^*_2}_0 C_{a1}(t)dt >0.
\end{equation}
\noindent
Note that
\begin{eqnarray} \label{equ:equnew}
& & -C_r + \lambda T{}^*_2 C_{a2}(T{}^*_2)-\lambda \int^{T{}^*_2}_0  
C_{a2}(t)dt 
\nonumber \\
&=& -C_r +\lambda T^*_2 C_{a1}(T^*_2)+\lambda T^*_2\Delta (T^*_2)-
\lambda \int ^{T^*_2}_0 C_{a1}(t)dt+\lambda \int ^{T^*_2}_0 
\Delta (t)dt \nonumber \\
&=& \left \{ -C_r +\lambda T^*_2 C_{a1}(T^*_2)-\lambda 
\int^{T{}^*_2}_0 C_{a1}(t)dt \right \} + 
\lambda \left \{ T^*_2 \Delta (T^*_2) - \int^{T^*_2}_0 
\Delta (t)dt \right \} \nonumber \\
& = &  \varphi (T^*_2 ) + \lambda \int^{T^*_2 }_0 
\left( \Delta (T^*_2 )-\Delta (t)\right ) dt \nonumber \\
& \geq & \varphi (T^*_2 )>0,
\end{eqnarray} 
\noindent
where Eq(\ref{equ:equnew}) comes directly
from the assumption that $\Delta (t)$ is 
nondecreasing in $t\geq 0$.  
However, as the optimal refresh policy  associated with $C_{a2}(t)$, 
it must be true that
$$-C_r+\lambda T^*_2 C_{a2}(T^*_2 )-\lambda \int^{T^*_2 }_0
C_{a2}(t)dt=0$$
\noindent
which contradicts with the inequality (\ref{equ:equnew}).  
Therefore, we must have $T^*_1 \geq T^*_2 $. 
\end{proof}

In the previous discussion 
we assume that 
refresh interval has a fixed length of $T$.
This assumption is somewhat unrealistic in practice. 
In the following, 
we study the impact of random refresh scheduling on 
the long-run mean average cost. 

\noindent Let $\{Y_i, i\geq 1\}$ be a sequence of {\it iid} random variables 
defined on $(0,\infty)$ following certain distribution $H$.
The sequence $\{Y_i, i\geq 1\}$ 
representing the interarrival times under a random refresh scheduling 
is assumed to be
independent of the Poisson arrival of content update at a server.
It is obvious that a fixed refresh policy is only a 
special case of a random refresh policy.  \\

Let ${\cal H}$ be the family of all 
distribution functions on $(0,\infty)$ with 
finite first moment. Namely,
$${\cal H}=\left \{H \colon ~H \hbox{~is~a~CDF ~on~$(0,\infty)$~and~}
\int^\infty _0\bar{H}(t)dt <\infty \right \}$$
\noindent
where $\bar{H}(t)\equiv 1-H(t),~\forall t\geq 0$.  

\begin{theorem} \label{theorem:the5}
Let $C_H(\cdot)$ denote the long-run mean average cost under 
a random refresh interval $Y$ characterized by
certain distribution function $H \in {\cal H}$,
and $C(T)$ denote the long-run mean average cost under fixed 
refresh scheduling with interval $T$,  then 
$$\min\limits _{H\in {\cal{H}}} C_H = \min\limits _{T>0} C(T).$$
\end{theorem}

\begin{proof}
Since the sequence $\{Y_i, i\geq 1 \}$ of 
interarrival refresh 
times is independent of the 
Poisson arrival of content update, 
it is easy to see 
that the random costs over the intervals 
$(0, Y_1],(Y_1, Y_1+Y_2], \ldots$ are {\it iid}. 
Using the same line as the proof 
of Theorem~\ref{theo:relationship},   
the long-run mean average cost is expressed as
$$C_H =\frac{E(\hbox{random~age-related~cost~over~}(0,Y])}{E(Y)},$$
\noindent
where $Y$ has distribution $H$. 
Let $\xi (Y)$ be the random cost in the cycle $(0, Y]$,
we have
\begin{eqnarray} \label {equ:equ30}
E ( \xi (Y))&=& E\{ E [\xi (Y)|Y]\}\nonumber \\
&=& \int ^\infty _0 E(C_r+\sum\limits ^{N(y)}_{n=1}C_a(y-S_n)I_{\{N(y)>0\}})dH(y), 
\end{eqnarray}
\noindent
where $S_n=\sum\limits ^n_{i=1}X_i$ 
denotes the time of the {\it n}th arrival of content update at the server. 
Due to the independence of  
$\{X_i, i\geq 1\}$ and $\{Y_i, i\geq 1\}$, 
from Eq(\ref{equ:equ30}) we further obtain 
\begin{eqnarray} \label{equ:equ31}
E(\xi (Y)) &=& \int ^\infty _0 (C_r + \int ^y_0 \lambda C_a(t)dt)dH(y) \nonumber \\
&=& C_r + \int^\infty _0 \left (\int ^y_0 \lambda C_a(t)dt \right ) dH(y).\nonumber
\end{eqnarray}
\noindent
Therefore, the long-run expected average cost is expressed as 
\begin{equation} \label{equ:equ32}
C_H=\frac{C_r}{E(Y)}+\frac{ \int^\infty _0 \left ( \int ^y_0 \lambda C_a(t)dt \right)
dH(y)}{E(Y)}.
\end{equation}
It is straightforward that 
\begin{equation}\label{equ:easy}
\min\limits_{H\in \cal{H}}C_H\leq \min\limits _{T>0}C(T)
\end{equation} 
since a fixed refresh scheduling with the interval $T>0$ is a degenerate 
random refresh scheduling when $P(Y=T)=1$.

Now for any given random refresh scheduling $Y$ with distribution $H\in {\cal{H}}$.
Eq(\ref{equ:equ32}) can be reexpressed as
\begin{eqnarray} \label{equ:equ33}
C_H&=& \frac{C_r}{E(Y)} + \frac{\int^\infty _0 \left 
( \int^\infty_t \lambda C_a(t)dH(y) \right)dt}{E(Y)} \nonumber \\
&=& \frac{C_r}{E(Y)} + \frac{\lambda \int ^\infty _0 C_a(t)\bar{H}(t)dt}{E(Y)}.
\end{eqnarray}
\noindent
If we choose $T= E(Y)\equiv \mu$, meaning that 
the fixed refresh interval $T$ equals to 
the mean value of the
random refresh interval $Y$,  
then according to Eq(\ref{equ:equc}), 
we have
\begin{equation} \label{equ:equ34}
C(T)= \frac{C_r}{\mu} + \frac{\lambda \int ^{\mu}_0 C_a(t)dt}{\mu}.
\end{equation}
\noindent
Subtracting Eq(\ref{equ:equ34}) from Eq(\ref{equ:equ33}), 
we obtain
\begin{eqnarray} \label{equ:equ35}
C_H-C(T)&=& \frac{\lambda}{\mu} 
\left \{ \int ^\infty _0 C_a(t)\bar{H}(t)dt - 
\int ^\mu_0 C_a(t)dt \right \} \nonumber \\
&=& \frac{\lambda}{\mu} \left \{ \int ^\mu _0 
C_a(t)\bar{H}(t)dt +  \int ^\infty _\mu C_a(t)\bar{H}(t)dt - 
\int ^\mu_0 C_a(t)dt \right \} \nonumber \\
&=& \frac{\lambda}{\mu} \left \{ \int ^\infty _\mu C_a(t)\bar{H}(t)dt -  
\int ^\mu _0 C_a(t)H(t)dt \right \} \nonumber \\
&\geq & \frac{\lambda}{\mu} \left 
\{ C_a(\mu) \int ^\infty _\mu \bar{H}(t)dt - 
C_a(\mu) \int ^\mu _0 H(t) dt\right \}   \nonumber \\
&=& \frac {\lambda C_a(\mu)}{\mu} \left \{ \int ^\infty _\mu \bar{H}(t)dt -
\int ^\mu _0 (1-\bar{H}(t))dt \right \} \nonumber \\
&=&   \frac {\lambda C_a(\mu)}{\mu} \left \{ \int ^\infty _0 
\bar{H}(t)dt - \mu \right \} =0  
\end{eqnarray}
\noindent thus we have 
\begin{equation}\label{equ:tt}
C_H \geq C(T)
\end{equation}
\noindent
Combining Eq(\ref{equ:easy}) and Eq(\ref{equ:tt}), 
we obtain
$$\min\limits _{H\in{\cal{H}}} C_H = \min\limits _{T>0} C(T)$$ 
\end{proof}

Such a mathematical equivalence between fixed refresh and 
random refresh schedulings offers a theoretical justification for  
the flexibility that is needed in real application environments.
Refresh scheduling with random variable $Y$ over a finite
interval $(a,b)$ is more flexible than fixed refresh scheduling.
Let $T>0$ be a given fixed refresh interval 
and $Y$ be a positive random interval 
following 
a distribution on interval $(a,b)$ with mean $E(Y)=T$.
From Eq(\ref{equ:tt}) it can be seen
that $C_H \geq C(T)$ and the equality holds if and only
if $Y$ is a degenerate random variable when
$P(Y=T)=1$. Thus,
for any given $T>0, \delta>0$ satisfying 
$T-\delta>0$, if
we denote ${\cal H}^* = \{ \mbox{H : H is a CDF on } 
(T-\delta,T+\delta) \mbox{ and } 
\int_{T-\delta}^{T+\delta} t dH(t) = T \}$,
then 
we have $C_H > C(T)$ for any non-degenerate $H \in {\cal H}^*$.
In particular, if $U(\delta)$ is the uniform distribution on 
$(T-\delta, T+\delta)$, then it is true that 
$C_{U(\delta)} > C(T)$. It is also easy to show that
$\lim_{\delta \rightarrow 0} C_{U(\delta)} = C(T)$. 

\section{Synchronizing Multiple Content Elements}
In the section we will extend this approach to
the case that involves more than one elements with different content 
update rates.

Consider a server containing content elements $S=\{e_1,\cdots, e_M\}$. 
Each element $e_i$ is being updated according to a Poisson process 
with intensity $\lambda_i$ ($1\leq i \leq M$). 
It is theoretically appealing to synchronize each content element 
$e_i$ using refresh interval $T_i$, which is referred to as 
{\it non-uniform allocation policy} \cite{Cho99,Cho2000}. 
However, 
{\it uniform allocation policy} \cite{Cho99,Cho2000}, i.e., 
synchronizing all content elements by using the same refresh 
interval $T$, is practically preferable by 
amortized cost analysis. The underlying reason is that
each refresh requires a connection establishment along with 
bandwidth usage and processing 
overhead. A connection overhead (latency) denoted as $C_{conn}$
is considered as the dominating factor 
in determining refresh cost $C_r$. 
The advantage of the {\it uniform allocation policy} 
over the {\it non-uniform allocation policy} is its ability to share its 
connection cost over the number of the content elements.  As
a result, the amortized 
connection for the {\it uniform allocation policy} is calculated as $C_{conn}/M$,
which is much cheaper than that of the 
{\it non-uniform allocation policy}, $C_{conn}$.
 
Let $\lambda_i$ be the intensity rate of Poisson process describing 
the content update of the cache element $e_i$, $ 1\leq i \leq M$.
From Theorem~\ref{theo:relationship}, 
the long-run mean average cost of caching over the entire  
$S$ is
$$ 
C(T)= \sum_{i=1}^M \bigl [\frac{C_r(\lambda_i)}{T} + 
\frac {\lambda_i \int^T_0  C_a(t; \lambda_i)dt}{T}\bigr ],$$
where $C_r(\lambda_i)$ is the cost associated with a refresh of element $e_i$, 
and $C_a(t; \lambda_i)$ is
the age-related cost function associated with 
element $e_i$, if the {\it uniform allocation policy} 
$T$ is applied.  Then
$C(T)$ can be rewritten as
\begin{equation}\label{equ:ppp}
C(T)= M [\frac{\bar{C}_r}{T} + \frac {\int^T_0  \bar{C}_a(t)dt}{T}],
\end{equation} 
where
\begin{equation}\label{equ:pppre}
\bar{C}_r=\frac{\sum_{i=1}^M C_r(\lambda_i)}{M}
\end{equation}
and
\begin{equation}\label{equ:pppa}
\bar{C}_a(t)=\frac{\sum_{i=1}^M \lambda_iC_a(t; \lambda_i)}{M}.
\end{equation}

\noindent If the size of the set of cache elements, 
$M$, is sufficiently 
large, 
then it is convenient to describe all $\lambda_i$ by a 
probability density function $g(\lambda)$ on $(0, \infty)$. 
With this assumption, the long-run mean average cost of caching 
the entire set of elements $S$ is given as
\begin{equation}\label{equ:ppp1} 
C(T)= M [\frac{\bar{C}_r}{T} + \frac {\int^T_0  \bar{C}_a(t)dt}{T}],
\end{equation}
where 
$$\bar{C}_r=\int_0^\infty C_r(\lambda)g(\lambda)\,d\lambda$$
and
$$\bar{C}_a(t)=\int_0^\infty \lambda C_a(t; \lambda)g(\lambda)\,d\lambda.$$

\noindent Combining Eq(\ref{equ:ppp}) and Eq(\ref{equ:ppp1}), 
we see that minimizing $C(T)$ is equivalent to minimizing
\begin{equation}\label{equ:ppp2}
\frac{\bar{C}_r}{T} + \frac {\int^T_0  \bar{C}_a(t)dt}{T}.
\end{equation}
\noindent 
Since the form of Eq(\ref{equ:ppp2}) is the same as Eq(\ref{equ:equc}),
we immediately obtain the following result.
\begin{theorem} 
Suppose that the arrival of update of each content $e$ 
is a Poisson process with intensity 
rate $\lambda$, $C_r(\lambda)$ is the cost associated 
with a refresh, and the refresh interval is $T$.  
Then the long-run mean average cost of caching $S$ is given as 
$$C(T)= M\bigl [\frac{\bar{C}_r}{T} + \frac {\int^T_0  \bar{C}_a(t)dt}{T}\bigr ],$$ 
where
$$\bar{C}_r=\int_0^\infty C_r(\lambda)\,dG(\lambda),$$
$$\bar{C}_a(t)=\int_0^\infty \lambda C_a(t; \lambda)\,dG(\lambda),$$
and $G(\lambda)$ is a distribution function on 
$(0, \infty)$ satisfying $\int_0^\infty \lambda C_a(t; \lambda)\,dG(\lambda)<\infty$ for any $t>0$. 
The problem of minimizing $C(T)$ is equivalent to minimizing $\bar{C}(T)$ given as
\begin{equation}
\bar{C}(T)=\frac{\bar{C}_r}{T} + 
\frac {\int^T_0  \bar{C}_a(t)dt}{T}.
\end{equation}
If 
the age-related cost function $\bar{C}_a(t)$ satisfies 
(a) $\bar{C}^{\prime}_a(t) >0$; (b)
$\bar{C}_a(\infty) = \infty$, then 
there exists unique refresh interval, designated $T^*$, which minimizes 
$C(T)$ and $\bar{C}(T)$. Moreover, $T^*$ is determined by 
the equation
$$T\bar{C}_a(T)-\int\limits^T_0 \bar{C}_a(t)dt=\bar{C}_r.$$
\end{theorem}

\section{Conclusion}
There is a fundamental tradeoff between the freshness of content and 
the overhead of refresh synchronization.  
An excessive refresh puts 
a strain on computation resource of network bandwidth and servers,
while a deficient refresh sacrifices the required freshness of content. 
This paper is focused on studying the effect of refresh scheduling 
on the freshness-cost tradeoff. 
We formulate the refresh scheduling problem 
with a generic cost model to capture the relationship between 
the arrival rate of content update, the freshness of content and the 
cost of refresh synchronization, and 
then use this cost model to  
determine an optimal refresh interval that minimizes the overall 
cost involved.

Theoretical results obtained in this paper
suggest that the optimal refresh 
frequency should be a function of 
the arrival rate of content update and the refresh cost,
implying that refresh frequency
is determined by the arrival rate of 
content update in order to
maintain a certain level of content freshness. 
Such a viewpoint has been implicitly reflected in the fact 
that many crawler-based Web search engines like {\it google}, {\it Inktomi} and 
{\it fast} started 
developing an automated tool to
identify the content change rate at some Web sites, and 
adapt refresh rate to 
the actual changing rate of content sources \cite{Notess2001}.
This paper gives a quantitative analysis of 
the freshness-cost tradeoff and 
offers a theoretical guidance in determining 
the best tradeoff between the freshness of content and 
the cost of refresh synchronization.

\section{Acknowledgement}
The authors would like to thank anonymous referees for their insightful comments
on a preliminary version of this paper.

\bibliography{cache}
\bibliographystyle{plain}

\end{document}